\documentclass[letter]{aa} % for the letters 
\usepackage{graphicx}
\usepackage{amsmath}
\usepackage{txfonts}
\begin{document}

   \title{Indications for an influence of hot Jupiters\\ on the rotation and activity of their host stars}
   
   \titlerunning{Hot Jupiters and stellar rotation and activity}
   
   \author{K.\ Poppenhaeger
          \inst{1,2}
          \and
          S.J.\ Wolk\inst{1}}

   \institute{\inst{1} Harvard-Smithsonian Center for Astrophysics,
              60 Garden St., Cambridge, 02138 MA, USA\\
              \inst{2} NASA Sagan Fellow\\
              \email{kpoppenhaeger@cfa.harvard.edu}
             }

   \date{Received January 17 2014; accepted March 27 2014}

% \abstract{}{}{}{}{} 
% 5 {} token are mandatory
 
  \abstract
   {The magnetic activity of planet-hosting stars is an important factor for estimating the atmospheric stability of close-in exoplanets and the age of their host stars. It has long been speculated that close-in exoplanets can influence the stellar activity level. However, testing for tidal or magnetic interaction effects in samples of planet-hosting stars is difficult because stellar activity hinders exoplanet detection, so that stellar samples with detected exoplanets show a bias toward low activity for small exoplanets.}
   {We aim to test whether exoplanets in close orbits influence the stellar rotation and magnetic activity of their host stars.}
   {We developed a novel approach to test for systematic activity-enhancements in planet-hosting stars. We use wide (several 100 AU) binary systems in which one of the stellar components is known to have an exoplanet, while the second stellar component does not have a detected planet and therefore acts as a negative control. We use the stellar coronal X-ray emission as an observational proxy for magnetic activity and analyze observations performed with Chandra and XMM-Newton.}
   {We find that in two systems for which strong tidal interaction can be expected the planet-hosting primary displays a much higher magnetic activity level than the planet-free secondary. In three systems for which weaker tidal interaction can be expected the activity levels of the two stellar components agree with each other.}
   {Our observations indicate that the presence of Hot Jupiters may inhibit the spin-down of host stars with thick outer convective layers. Possible causes for this effect include a transfer of angular momentum from the planetary orbit to the stellar rotation through tidal interaction, or differences during the early evolution of the system, where the host star may decouple from the protoplanetary disk early because of a gap opened by the forming Hot Jupiter.}


   \maketitle
%
%________________________________________________________________

\section{Introduction}

Stellar magnetic activity is a phenomenon that shapes the environment of exoplanets. Magnetically induced processes such as stellar flares, high-energy emission, and coronal mass ejections can have profound effects on the atmospheres of close-in exoplanets, causing the heating of high-altitude layers and atmospheric evaporation \citep{Vidal-Madjar2003, Murray-Clay2009, Lammer2003, Khodachenko2012}.

It is well known that the activity of cool stars decreases over time; because rotation is the driver of activity, the magnetic braking caused by the stellar wind causes all activity processes to weaken over timescales of gigayears. Some processes can preserve high activity levels over long timescales, such as tidal locking in close binaries, which sustains fast stellar rotation and therefore activity. It has been speculated that similar effects, albeit on a weaker scale, may occur for stars with Hot Jupiters \citep{Cuntz2000, Lanza2008}. Observational studies have been performed on individual systems \citep{Shkolnik2005, Shkolnik2008, PoppenhaegerLenz2011, Pillitteri2011, Miller2012} as well as on larger samples of planet-hosting stars \citep{Kashyap2008, Poppenhaeger2010, Shkolnik2013}, finding weak correlations of stellar activity with the presence of Hot Jupiters. Indications for higher $v\sin i$ values have also been reported \citep{Pont2009tidal} for systems with Hot Jupiters compared with systems hosting smaller or more distant planets, yielding first indications of a tidal influence of exoplanets on their host stars. This was solidified in a detailed study by \cite{Husnoo2012}, who found indications for excess rotation of several hot Jupiter host stars. However, unambiguous signatures of a planet-induced enhanced activity level are difficult to identify, because planet-detection methods favor magnetically inactive stars. Active stars usually only allow for the detection of planets with strong RV signatures or deep and frequent transits (i.e.\ Hot Jupiters). This bias induces spurious trends in the population of stars with detected planets \citep{Poppenhaeger2011}. 

We have developed an observational approach to test for planetary influences on the stellar activity level without most of the common biases induced by planet detection. We have selected a small sample of planet-hosting stars that have a known stellar companion; the stellar companions are not known to have planets themselves. Companionship of the two stellar components has been established through common proper motion or common radial velocity; the two stars can therefore be assumed to have the same age. The distance between these stellar components is large enough so that no influence on the activity level can be expected. For close (<0.1 AU) or medium-distance (<10 AU) binaries, such an influence has been observed and traced back to tidal locking or, in the case of moderate distances, to differences in the circumstellar disk evolution \citep{Meibom2007, Morgan2012}. Our systems, however, have binary distances of $>100$ AU, for which such trends are absent. We list our sample of five such systems in Table~\ref{systems}. In these systems, the stellar companions without a known planet act as a negative control group to the planet-hosting stars whose activity level may have been influenced by their planets.

\begin{table*}
\renewcommand{\tabcolsep}{0.07cm}
\begin{footnotesize}
\begin{tabular}{l c c c c c c c c c c c c c c } \hline \hline
Star		& spec.\ type	& B-V	& ang.\ sep.\	& dist.	& $P^{\ast}_{rot}$	& $M_P$ & $P_{orb}$	& $a_{sem}$	&$\log R^\prime_{HK}$	& $\log L_X$	& apparent age	& $J_{\ast}$	& $J_{orb}$ & $h_{tide}/h_{scale}$\\ 
		&		&	& (arcsec)	& (pc)	& (d)			& ($M_{Jup}$)& (d)	& (AU)		&			& (erg\,s$^{-1}$)& 		& (kg m$^2$ d$^{-1}$)	& (kg m$^2$ d$^{-1}$)& \\ \hline
\object{HD\,189733}\,A	& K0V		&0.93	& 11.2''	& 19.3	& 12			& 1.14	& 2.22		& 0.031		&-4.501	& 28.2 & 1-2 Gyr	& 9.3e46&1.3e47	&0.008\\
HD\,189733\,B	& M4V		&	& 11.2''	& 19.3	& 			& 	& 		& 		&	& 26.7 & $\geq$ 5 Gyr	&	&	&\\
\object{CoRoT-2}\,A	& G7V		&0.854	& 4''		& 270	& 4.5			& 3.31	& 1.74		& 0.028		&-4.331	& 29.2 & 0.1-0.3 Gyr	&4.3e47	&4.0e47	&0.072\\
CoRoT-2\,B	& K9V		&	& 4''		& 270	& 			& 	& 		& 		&	& 27.0 & $\geq$ 5 Gyr 	&	&	&\\
\object{$\tau$ Boo}\,A	& F7V		&0.52	& 2.5''		& 15.0	& 3.3			& 8.14	& 3.31		& 0.048		&-4.70	& 28.8 & 1-2 Gyr	&2.0e48	&1.5e48	&  0.047\\
$\tau$ Boo\,B	& M		&	& 2.5''		& 15.0	& 			& 	& 		& 		&	& 27.65 & 1-3 Gyr	&	&	&\\
\object{$\upsilon$ And}\,A& F8IV-V	&0.53	& 52''		& 13.5	& 9.5			& 0.62	& 4.62		& 0.059		&-5.04	& 27.7 & $\geq$ 5 Gyr	&6.4e47	&1.4e47	&0.002\\
$\upsilon$ And\,B&M4V		&	& 52''		& 13.5	& 			& 	& 		& 		&	& 26.45 & $\geq$ 5 Gyr	&	&	&\\
\object{55\,Cnc}\,A	& G8V		&0.87	& 81''		& 13	& 43			& 0.026	& 0.74		& 0.015		&-5.04	& 27.07 & $\geq$ 5 Gyr	&4.6e46	&2.3e45	&0.005\\
55\,Cnc\,B	& M3-4V		&	& 81''		& 13	& 			& 	& 		& 		&	& 26.22 & $\geq$ 5 Gyr	&	&	&\\
\hline\hline
\end{tabular}
\end{footnotesize}
\caption{Wide binary systems consisting of two main-sequence cool stars with at least one exoplanet for which X-ray data has been collected.  The closest exoplanet is given in case of multiple exoplanets per system. 
}
\label{systems}
% \vspace{-0.6cm}
\end{table*}

\section{Data analysis}

Stellar activity and rotation can be measured by several observational proxies. The most common ones are coronal X-ray emission, chromospheric emission in the Ca II H and K lines around 3950\,\AA, and direct measurements of the stellar rotation period through broadband light-curve modulations. The projected stellar rotation velocity $v\sin i$ is easy to observe, but is an ambiguous measurement of stellar rotation if the stellar inclination is unknown. The planet-hosting stars in our sample, which are the primaries in their respective system, are well studied in the literature. The secondaries, however, are rather faint M or late-K dwarfs. Measurements of their rotational period or the blue part of their optical spectra that contains the Ca II H and K lines are therefore challenging, and we have therefore chosen to obtain X-ray observations of the primaries and secondaries to measure their magnetic activity level.

These observations were performed with {\it XMM-Newton} and {\it Chandra}. We present three new observations of the systems $\tau$~Boo AB, $\upsilon$ And AB, and 55 Cnc AB; the observations of HD~189733 AB and CoRoT-2 AB have been analyzed before \citep{Poppenhaeger2013, Schroeter2011} and are summarized here. To derive the stellar X-ray luminosities, we extracted the detected source photons and compared them with the background signal from a source-free nearby region. We then converted the detected background-subtracted source count rate into a source flux using WebPIMMS. We checked that all newly observed sources display a soft coronal spectrum with a hardness ratio $(H-S)/(H+S) < 0$, using an energy range of $0.2-1$\,keV for the soft band ($S$) and $1-3$\,keV for the hard band ($H$). We therefore chose to adopt a mean coronal temperature of 3\,MK and 5\,MK for the F/G/K and M dwarfs to derive their X-ray fluxes.

\begin{table}
\renewcommand{\tabcolsep}{0.2cm}
\begin{footnotesize}
\begin{tabular}{l c c c c c c c c} \hline \hline
System & obs.\ with		& ObsID		& obs.\ date	& exp.\ time		\\ \hline
$\tau$ Boo AB & C	& 13232		& 2011-08-28	& 5.0 ks	 \\
$\tau$ Boo AB & C	& 13233		& 2012-04-21	& 4.9 ks	  \\
55 Cnc AB & X	& 0551020801	& 2009-04-11	& 13.9 ks	 \\
$\upsilon$ And AB & X	& 0722030101	& 2013-08-04	& 17.0 ks	 \\
\hline
\hline
\end{tabular}
\end{footnotesize}
\caption{New X-ray observations of the systems. C: {\it Chandra} ACIS-I, X: {\it XMM-Newton} EPIC.}
\label{obs}
% \vspace{-0.6cm}
\end{table}

%
%                                                One column figure
%----------------------------------------------------------- S_vib
   \begin{figure*}
   \centering
   \includegraphics[width=0.25\textwidth]{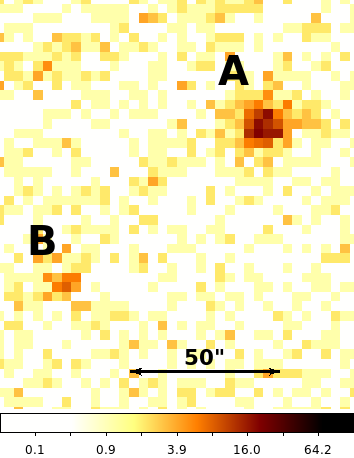}
   \hspace{0.8cm}
   \includegraphics[width=0.25\textwidth]{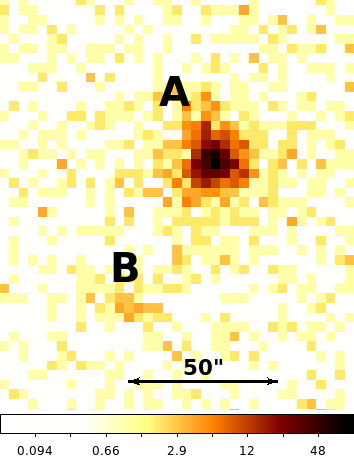}
   \hspace{0.8cm}
   \includegraphics[width=0.25\textwidth]{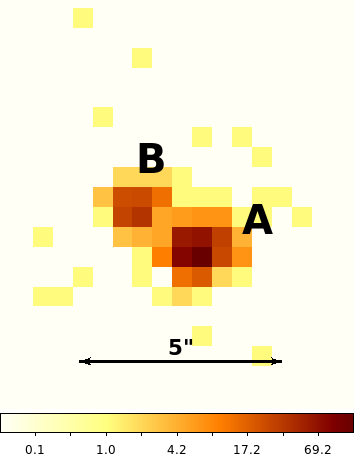}
   
      \caption{Left to right: X-ray images of 55 Cnc AB ({\it XMM-Newton} EPIC), $\upsilon$ And AB ({\it XMM-Newton} MOS1$+$MOS2), and $\tau$ Boo AB ({\it Chandra} ACIS-I), energy band 0.2-3 keV.}
         \label{lx}
   \end{figure*}
%
%______________________________________________________________

\section{Results}

All stars were individually detected in X-rays, with the exception of CoRoT-2B, for which a strong upper limit of $\log L_X < 27.0$\,erg\,s$^{-1}$ was derived by \cite{Schroeter2011}. All sources display a low-level varibility in their light curves with at most a change of a factor of two in count rate. Strong X-ray flares are therefore absent\footnote{Small flares with count rate increases of $\leq 2$ were observed for HD 189733 A by \cite{Pillitteri2011} and \cite{Poppenhaeger2013}; we will refer to the quiescent luminosities here.}, because they often display count rate increases of a factor of 30 or more \citep[for example]{Schmitt2008CNLeo, Fuhrmeister2011}.

To derive the apparent ages of the stellar components we compared their X-ray emission with X-ray observations of stellar samples of different ages, namely individual WTTS (10 Myr), stars in the Pleiades (85 Myr), and stars in the Hyades (650 Myr) from \cite{Stelzer2001}, field stars (about 5 Gyr) from \cite{Schmitt2004}, and old disk and halo M dwarfs ($\geq$ 10 Gyr) from \cite{Micela1997}. We used the median X-ray luminosities and standard deviations from these authors; if no standard deviation was given, we fitted the measured luminosities with a log-normal distribution to derive the median and the standard deviation. The literature gives values for the chromospheric Ca~II H and K emission and the stellar rotation period for most of the primaries; for the chromospheric data, we used the age calibration given by \cite{Mamajek2008}, while we compared the rotational periods with rotation measurements in various stellar clusters \citep{Meibom2011_1Gyr, Meibom2007}.

{\bf $\upsilon$ And AB:} $\upsilon$ And A is an inactive F star. Its X-ray luminosity was measured multiple times in 2009 with an almost constant level of $\log L_X = 27.6$ to $27.8$\,erg\,s$^{-1}$. In our {\it XMM-Newton} observation from 2013, it displays a background-subtracted count rate of 0.023 cps in the combined MOS detectors, corresponding to an X-ray luminosity of $\log L_X = 27.75$\,erg\,s$^{-1}$. This is slightly lower than the typical X-ray luminosities found for main-sequence field stars of spectral type F and G, ($\log L_X = 28.05 [27.12, 28.98]$\,erg\,s$^{-1}$, \citet{Schmitt2004}). Its X-ray emission therefore suggests an age of about 5 Gyr or older. This is consistent with the observation that $\upsilon$ And A has very likely started to evolve off the main sequence: its absolute magnitude, $M_V = 3.44$, is about 0.6 mag brighter than expected if the star were on the main sequence, $M_{V,\,exp}=4.02$, using the main-sequence description given by \cite{Wright2004maunder, Wright2005maunder}. With a stellar mass of $1.31 M_\odot$, its main-sequence life time is about 5 Gyr, so the star is presumably older than that. This is consistent with its low chromospheric Ca II H and K line emission of $\log R^\prime_{HK} = -5.04$ \citep{Wright2004chrom}.

% 200-3000 eV
% log T = 6.5
% thick filter
% ups And A
% mean mos count rate (15''): (827. - 429./(60./15.)**2) / 17400. / 2. = 0.02299 cps
% -> 1./0.7 * 0.02299 = 0.03284 cps is actual count rate
% -> 2.758E-13 ergs/cm/cm/s -> log LX = np.log10(4*np.pi*(3e18*13.5)**2 * 2.758E-13) = 27.75

$\upsilon$ And B is an M dwarf companion to $\upsilon$ And A. Here we report its detection in soft X-rays. It was observed for 17 ks with {\it XMM-Newton} in 2013; the source was too close to the detector edge in the PN detector to include the PN data in the analysis, so we used the signal from the MOS detectors alone. $\upsilon$ And B displayed a background-subtracted count rate of 0.003\,cps in the energy band 0.2-3.0\,keV in the two MOS detectors combined, corresponding to an X-ray luminosity of $\log L_X = 26.45$\,erg\,s$^{-1}$. This agrees well with the X-ray luminosities observed for M dwarfs in the Galactic old disk and halo population, $\log L_X = 26.49 [25.76, 27.21]$\,erg\,s$^{-1}$ \citep{Micela1997}. $\upsilon$ And B is therefore most likely an old M dwarf with an age $> 5$\,Gyr, consistent with that of the primary.

% 200-3000 eV
% log T = 6.7
% thick filter
% ups And B
% mean mos count rate (one detector) (15''): (72. - 429./(60./15.)**2) / 17400. / 2. = 0.001298 cps
% -> 1./0.7 * 0.001298 = 0.001854 cps is actual count rate
% -> 1.363E-14 ergs/cm/cm/s -> log LX = np.log10(4*np.pi*(3e18*13.5)**2 * 1.363E-14) = 26.45
% s 0.2-1.0: 38 - 141./(60./15.)**2 = 29.19
% h 1.0-3.0: 34 - 288./(60./15.)**2 = 16.0

% # main sequence life time:
% t = (m_star/m_sun)**-2.5 * t_sun
% # abs. mag:
% -math.log((13.5/10)**2, 2.5) + 4.10
% # wright main sequence:
% bv = 0.53
% a = np.array([1.11255, 5.79062, -16.76829, 76.47777, -140.08488, 127.38044, -49.71805, -8.24265, 14.07945, -3.43155])
% ms = np.sum([bv**i*a[i] for i in np.arange(0,9)])

% def ms(a, bv):
%   ms = 0
%   for i in np.arange(0, len(a)):
%     ms = ms + bv**i * a[i]
%   
%   return ms
% 
% 
% bv = np.linspace(0, 1.5, 100)
% ms_list = np.zeros(len(bv))
% for i in np.arange(0, len(ms_list)):
%   ms_list[i] = ms(a, bv[i])
% 
% plt.plot(bv, ms_list)

{\bf 55 Cnc AB:} 55 Cnc A is an inactive K dwarf. Its X-ray luminosity has been measured with {\it XMM-Newton} to be $\log L_X = 27.07$\,erg\,s$^{-1}$ \citep{Poppenhaeger2010}. It is fainter in X-rays than typical field K dwarfs ($\log L_X = 27.71 [27.1, 28.32]$\,erg\,s$^{-1}$), indicating an old age of $\geq 5$\,Gyr. Its chromospheric Ca II H and K line emission is low as well with $\log R^\prime_{HK}=-5.04$ \citep{Wright2004chrom}, consistent with old age. Its rotation period of ca.\ 43\,d \citep{Fischer2008} matches these indications.

55 Cnc B was included in the field of view of 55\,Cnc\,A's 14\,ks X-ray observation, but has not been analyzed so far. Here we report its X-ray detection with a background-subtracted count rate of 0.005 in the 0.2-3\,keV energy band of the combined MOS and PN detectors. This corresponds to an X-ray luminosity of $\log L_X = 26.22$\,erg\,s$^{-1}$. Considering that its X-ray emission level is similar to halo M dwarfs ($\log L_X = 26.49 [25.76, 27.21]$\,erg\,s$^{-1}$), it indicates a very old age of $5-10$\,Gyr for this secondary, consistent with the primary's estimated age.

% 200-3000 eV
% log T = 6.7
% thick filter
% 55 Cnc B
% EPIC count rate (15''): (100. - 242./(40./15.)**2) / 13500. = 0.00489 cps
% PN count rate (15''): (84. - 320./(40./15.)**2) / 12660. = 0.00308 cps
% -> 1./0.67 * 0.00308 = 0.004597 cps is actual pn count rate
% -> 7.123E-15 ergs/cm/cm/s -> log LX = np.log10(4*np.pi*(3e18*13)**2 * 7.123E-15) = 26.13
% mean mos count rate (15''): (39. - 88./(40./15.)**2) / 13620. / 2. = 0.000977 cps
% -> 1./0.7 * 0.000977 = 0.001396 cps is actual count rate
% -> 1.025E-14 ergs/cm/cm/s -> log LX = np.log10(4*np.pi*(3e18*13)**2 * 1.025E-14) = 26.29
% s 0.2-1.0: 73 - 123./(40./15.)**2 = 55.70
% h 1.0-3.0: 27 - 120./(40./15.)**2 = 10.13

{\bf $\tau$ Boo AB:} $\tau$ Boo A is a late-F dwarf that has been shown to display different rotation, with a rotation period of $P_{rot}=3.0$\,d at the equator and $P_{rot}=3.9$\,d at the poles \citep{Donati2008}. Although magnetic field reconstructions indicate a magnetic cycle \citep{Fares2010}, its X-ray luminosity has been observed to be almost constant in the 0.2-3\,keV band between $\log L_X = 28.7$ and $28.9$\,erg\,s$^{-1}$ in 2003, 2010, and 2011 \citep{Poppenhaeger2012AN}. These observations did not resolve the M dwarf companion and therefore yielded the combined flux of both stars. But the contribution of the M dwarf is small, see below. This emission level is at the high end of observed field F and G dwarfs ($\log L_X = 28.05 [27.13, 28.98]$\,erg\,s$^{-1}$), but lower than G dwarfs in the Hyades ($\log L_X = 28.97 [28.91, 29.03]$\,erg\,s$^{-1}$), therefore suggesting an age of about 1\,Gyr. Its chromospheric Ca II H and K emission, $\log R^\prime_{HK}=-4.70$ \citep{Wright2004chrom}, yields a compatible age estimate of 1-2\,Gyr. Its rotation period of 3-4\,d also matches rotation periods found for stars of this spectral type in the 1 Gyr old cluster 6811 (2-6\,d) \citep{Meibom2011_1Gyr}. 

$\tau$ Boo B is a mid-M dwarf, located at an angular distance of 2.2$^{\prime\prime}$ from $\tau$ Boo A. We have observed $\tau$ Boo A and B with Chandra in two exposures of 5 ks each during 2012, and were able to detect the two components individually in X-rays. The observations were performed with ACIS-I, which is sensitive to energies between 0.6-15 keV. These observations therefore miss some soft X-ray flux between 0.2-0.6 keV, which is ususally detected with XMM-Newton or Chandra's ACIS-S. In the 0.6-5 keV band, we detect $\tau$ Boo B with an average X-ray luminosity of $\log L_X = 27.47$\,erg\,s$^{-1}$ and $\log L_X = 27.81$\,erg\,s$^{-1}$ in the two exposures, respectively. $\tau$ Boo A displayed an X-ray luminsity of $\log L_X = 28.46$  and $28.53$\,erg\,s$^{-1}$, respectively. The combined X-ray luminosity of the two stars in the 0.6-3\,keV is about a factor of two lower than in the previous full-band (0.2-3\,keV) observations performed with XMM-Newton. This is expected because $\tau$ Boo A's spectrum has been found to have a mean coronal temperature of $T=3$\,MK in the XMM-Newton observations, which corresponds to a flux reduction of about 50\% when moving to the harder X-ray band of Chandra's ACIS-I. M dwarfs often display higher mean coronal temperatures and a harder spectrum, so that $\tau$ Boo B is most likely missing less coronal flux in the ACIS-I observations. As two extremes, we assume that $\tau$ Boo B's true X-ray luminosity in the 0.2-3\,keV band is between $\log L_X = 27.5$ and $27.8$\,erg\,s$^{-1}$. This is at the high end of field M dwarfs ($\log L_X = 27.10 [26.43, 27.78]$\,erg\,s$^{-1}$), but lower than X-ray luminosities of M dwarfs in the Hyades ($log L_X = 27.95 [27.85, 28.05]$\,erg\,s$^{-1}$ ). Similar to $\tau$ Boo A, the most likely age for $\tau$ Boo B is therefore in the range of 1-3\,Gyr.

% 200-3000 eV (but eff. area starts at 600 eV)
% log T = 6.7
% tau Boo B
% ACIS-I count rate obs 1: 149./6384. = 0.0233 cps -> 
% 2.537E-13 ergs/cm/cm/s -> log LX = np.log10((15*3e18)**2 * 4*np.pi* 2.537E-13) = 27.81
% ACIS-I count rate obs 2: 81./7636. = 0.0106 cps -> 
% 1.154E-13 ergs/cm/cm/s -> log LX = np.log10((15*3e18)**2 * 4*np.pi* 1.154E-13) = 27.47

% log T = 6.5
% tau Boo A
% ACIS-I count rate obs 1: 457./6384. = 0.071585 cps -> 
% 1.333E-12 ergs/cm/cm/s -> log LX = np.log10((15*3e18)**2 * 4*np.pi* 1.333E-12) = 28.53
% ACIS-I count rate obs 2: 461./7636. = 0.0603719 cps -> 
% 1.124E-12 ergs/cm/cm/s -> log LX = np.log10((15*3e18)**2 * 4*np.pi* 1.124E-12) = 28.46

{\bf HD 189733 AB:} HD 189733 A has been observed with modern X-ray telescopes multiple times from 2007 to 2012, and its X-ray luminosity has been found to be fairly constant with $\log L_X = 28.05$ to $28.3$\,erg\,s$^{-1}$ \citep{Pillitteri2011, Poppenhaeger2013}. For a K star, this is at the high end of the X-ray luminosities observed for field stars, which display a median of $\log L_X = 27.71 [27.11, 28.32]$\,erg\,s$^{-1}$, as observed by \cite{Schmitt2004}. Younger single K stars in the Hyades, that is, with an age of ca.\ 650\,Myr, display median X-ray luminosities of  $\log L_X = 28.41 [28.26, 28.56]$\,erg\,s$^{-1}$ \citep{Stelzer2001}. This suggests an apparent age of about 1-2 Gyr for HD 189733 A. This agrees with its observed chromospheric emission in the Ca II H and K lines of $\log R^\prime_{HK}=-4.501$ \citep{Knutson2010}, which corresponds to an age of about 1-1.5\,Gyr. Moreover, its rotation period of 11.95\,d \citep{Henry2008} agrees well with the rotation periods observed for K stars in the 1 Gyr old open cluster NGC 6811, where typical rotation periods of 11\,d are found \citep{Meibom2011_1Gyr}. 

Rotation periods or chromospheric activity measurements are not available for HD 189733 B, but it has recently been detected in X-rays using Chandra observations with an X-ray luminosity of $\log L_X = 26.67$\,erg\,s$^{-1}$ \citep{Poppenhaeger2013}. Previous observations with XMM-Newton had yielded a consistent upper limit of $\log L_X < 26.9$\,erg\,s$^{-1}$ \citep{Pillitteri2010, Pillitteri2011}. Interestingly, this suggests an older age for this physical companion than the activity level of the primary indicates. Field M dwarfs display X-ray luminosities of $\log L_X = 27.11 [26.43, 27.78]$\,erg\,s$^{-1}$ \citep{Schmitt2004}, while M dwarfs of the Galactic old disk and halo population, that is, with ages of 10 Gyr and older, display luminosities of $\log L_X = 26.49 [25.76, 27.21]$\,erg\,s$^{-1}$ \citep{Micela1997}. The observed low X-ray luminosity points toward an age range of 5-10 Gyr for the secondary.

{\bf CoRoT-2 AB:} CoRoT-2 A is an active late-G dwarf. Its X-ray luminosity was observed by \cite{Schroeter2011} to be $\log L_X = 29.3$\,erg\,s$^{-1}$, which is compatible with an age of a few hundred Myr. G stars in the Pleiades typically have X-ray luminosities of $\log L_X = 28.98 [28.83, 29.13]$\,erg\,s$^{-1}$. CoRoT-2\,A displays strong emission in the chromospheric Ca II H and K lines with $\log R^\prime_{HK}=-4.331$ \citep{Knutson2010}, again indicating a young stellar age of about 200-300 Myr. Its rotation period is $P_{rot}=4.54$ \citep{Silva-Valio2011}, again confirming young age; similar rotation periods of about 5 days are found for stars of this spectral type in the 150\,Myr old cluster M35 \citep{Meibom2009_M35}. We therefore estimate CoRoT-2\,A's age to be between 100--300\,Myr.

CoRoT-2 B is a late K-dwarf that is not detected in the {\it Chandra} observation of CoRoT-2 A. \cite{Schroeter2011} derived an upper limit on its X-ray luminosity of $\log L_X < 26.9$\,erg\,s$^{-1}$. This is strongly at odds with the X-ray emission one would expect for a common young age of the primary and the secondary: young K dwarfs display typical X-ray luminosities around $\log L_X = 28.83$\,erg\,s$^{-1}$ (Pleiades, 85 Myr) and $\log L_X = 28.41$\,erg\,s$^{-1}$ (Hyades, 650 Myr). The actual X-ray luminosity of CoRoT-2 B is even lower than observed for field stars ($\log L_X = 27.71 [27.11, 28.32]$\,erg\,s$^{-1}$), which indicates an old age $>5$\,Gyr for this secondary.

%--------------------------------------------------------------
   \begin{figure}
   \centering
   \includegraphics[width=0.49\textwidth]{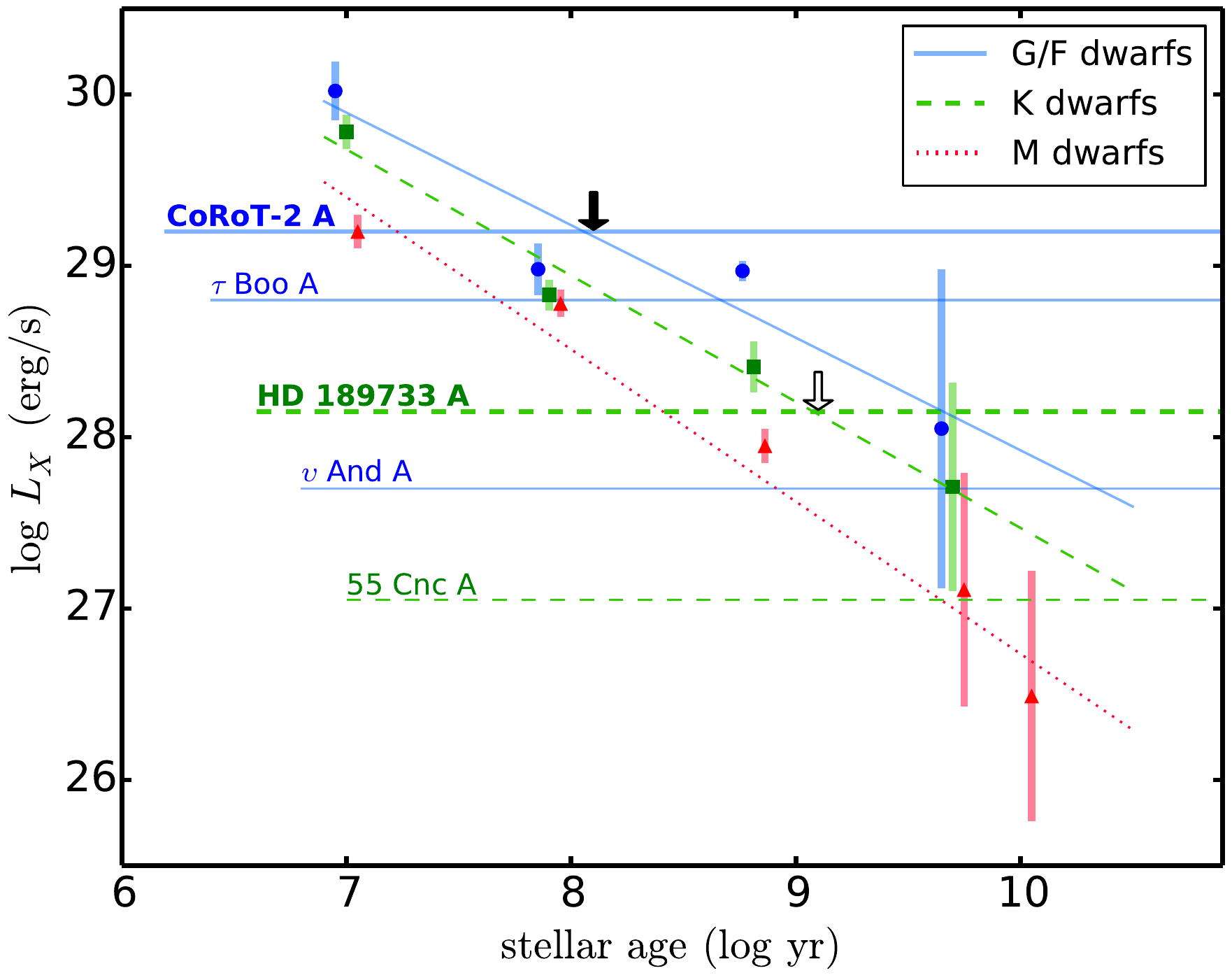}
   \includegraphics[width=0.49\textwidth]{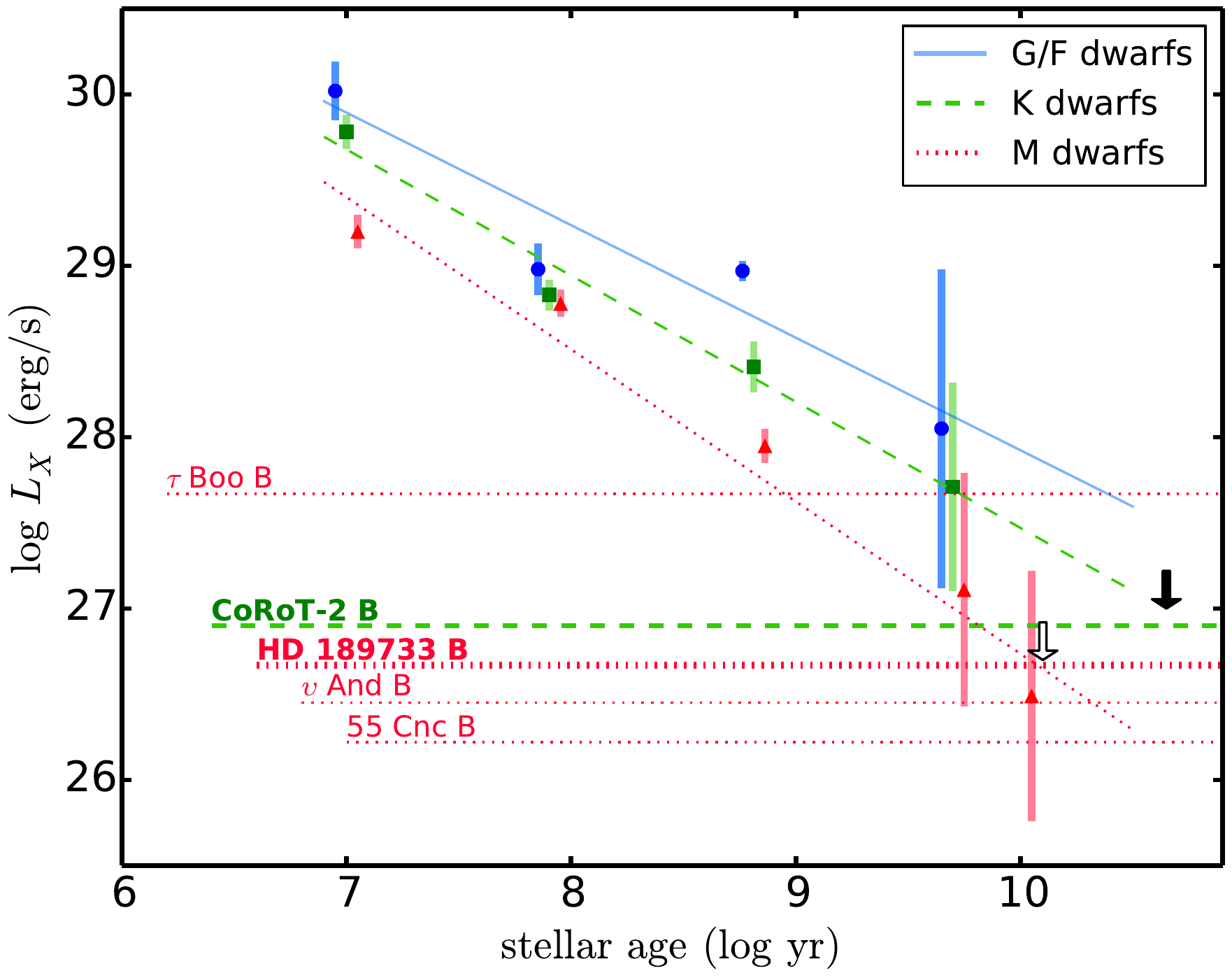}
      \caption{X-ray luminosity of the planet-hosting primaries (top) and the secondaries (bottom), shown as horizontal lines, in comparison with stellar populations with known ages (WTTS, Pleiades, Hyades, field stars, and old disk/halo stars). The pairs HD 189733 AB and CoRoT-2 AB are shown in bold face and with thicker lines; they display X-ray luminosities that are inconsistent with a common age for their A and B components (black arrows for CoRoT-2 AB, white arrows for HD 189733 AB).}
         \label{lx}
   \end{figure}
%______________________________________________________________

\section{Discussion}

In three of the observed systems, the age estimates for the primary (planet-hosting) star and the secondary agree well: in $\upsilon$ And AB and 55 Cnc AB, both stellar components seem to be of old age $\geq 5$\, Gyr, while in $\tau$ Boo AB both components appear to be about 1-3 Gyr old, that is, slightly younger than the field star population. However, in the two other systems, HD 189733 AB and CoRoT-2 AB, the apparent ages of the stellar components differ. For HD 189733 AB the difference is moderate, the primary's apparent age is somewhat younger than the field star population and the secondary is closer to the age of the old disk population. For CoRoT-2 AB the difference is very pronounced: CoRoT-2 A's apparent age is young with a few hundred Myr, and the secondary is very likely older than the field stars. 

Such a pronounced activity difference in physical companions is not common; \cite{Mamajek2008} found closely agreeing activity-age estimates for their physical companions when comparing the chromospheric Ca II H and K emission. The typical scatter in $\log R^\prime_{HK}$ is 0.07 dex, corresponding to roughly a factor of two in age. Stellar cycles cannot explain the observed behavior either, because most of the primaries in our sample have been observed multiple times without showing large changes in X-ray luminosity. A possible explanantion for the age discrepancies we observe for some of our targets may be an interaction with the extrasolar planet that orbits the primary star. The possibility of such interactions between planet and host star has been discussed by several authors; the two commonly discussed scenarios are magnetic and tidal interaction \citep{Cuntz2000}. In our two discrepant systems {\it both} the rotation and activity level of the planet-hosting primaries are higher than expected from the secondary, so that a transfer of angular momentum from the planetary orbit to the stellar rotation, induced by tidal interaction, could explain the observations. 

In this scenario, the planetary orbital period needs to be shorter than the stellar rotation period; furthermore, the potential energy of tidal bulges raised by the planet on the stellar surface would need to be dissipated efficiently to cause a spin-up or, as magnetic braking of the host star is occurring, an inhibited spin-down. Several authors have argued that thick convective outer layers probably provide a more efficient mechanism to dissipate this energy through convective eddies than a mostly radiative outer layer \citep{Albrecht2012, Zahn2008}; however, a realistic treatment of tidal interaction and stellar quality factors is challenging, because resonances occur for certain ratios of the orbital and spin periods of the bodies \citep{Ogilvie2007}. Recent theoretical models indicate that the relevant stellar quality factors for star-planet systems may be higher than thought before, on the order of $10^8-10^9$ \citep{Penev2011}, meaning that stars would dissipate the tidal energy from interactions with their planets very slowly.

For a rough comparison, we first take a look at the angular momentum of the stellar rotation and the planetary orbit in each system (if multiple planets exists, we take the planet on the closest orbit). As an order-of-magnitude estimate, we approximated the stellar angular momentum as that of a rotating solid sphere\footnote{For the Sun, the angular momentum was helioseismologically determined to be $1.9\times 10^{41}$\,kg\,m$^2$\,s$^{-1}$ \citep{Pijpers1998}, which is about 6\% larger than the estimate derived assuming a solid rotating sphere.}, $J_\ast = \frac{4}{5} \pi M_\ast R^2_\ast P^{-1}_{rot}$, and the angular momentum of the planetary orbit is given by $J_{orb} = 2 \pi M_p a^2_{sem} P^{-1}_{orb}$. We list the resulting values in Table~\ref{systems}; for the star-planet systems HD 189733 Ab, CoRoT-2 Ab, and $\tau$ Boo Ab, the angular momentum of the planetary orbit is larger than or similar to the angular momentum of the star. If the planet has migrated -- and is still migrating -- inward to the host star, a sufficient budget of angular momentum existed to inhibit the stellar spin-down from occurring at the typical rate. 

As a second comparison, we calculated the tidal bulge heights $h_{tide}$ raised on the stellar surface as a fraction of the scale height of the stellar photosphere $h_{scale}$, as described in \cite{Cuntz2000}. We list $h_{tide}/h_{scale}$ in Table~\ref{systems}. The relative bulge heights are largest for CoRoT-2 A and $\tau$ Boo A; however, $\tau$ Boo A, an F7V dwarf, has a very thin outer convection layer and may be much less efficient at dissipating the energy deposited in the tidal bulges. The next-largest relative bulge height is raised on HD 189733 A, the primary for which we observed a mild overactivity and overrotation compared with the secondary. We point out here that our findings that CoRoT-2~A and HD~189733~A are over-active agree well with the results of \cite{Husnoo2012}, who found these systems to be over-rotating compared to rotation isochrones. These authors identified four other systems with probable over-rotation: CoRoT-18, HAT-P-20, WASP-19, and WASP-43. Out of these, only HAT-P-20 is currently known to possess a stellar companion \citep{Bakos2011}, and we have arranged for follow-up observations of the stellar activity for this system.

Another scenario that might explain the apparent differences in stellar rotational evolution for the two systems concerns the rotational period with which stars enter the main sequence. Modern models of stellar spin-down have shown that the evolution of stellar angular momentum strongly depends on the initial stellar rotation \citep{Barnes2010}. Before a star enters the main sequence, its rotation is influenced by magnetic coupling to the inner circumstellar disk; after the disk resolves, the star  spins up because of its ongoing contraction, and then spins down because of magnetic breaking. A massive planet migrating inward opens a gap in the circumstellar disk; when it migrates into a close orbit around the host star, the gap it opens may change the star-disk coupling. If this has led to a different stellar rotation on the zero-age main sequence, these initial differences may have led to the over-spinning of HD 189733 A and CoRoT-2 A.

\section{Conclusions}

   \begin{enumerate}
      \item We presented X-ray luminosities of the individual stellar components in five wide binary-star systems in which one of the stars is known to host an exoplanet.
      \item We showed that for two of the systems -- those for which one expects the strongest tidal interaction between planet and host star -- the X-ray emission of the planet-hosting primary is stronger than expected from the secondary, assuming a common stellar age. These host stars are also over-rotating in comparison with the activity level of the secondary.
      \item We interpret this as an indication that a tidal influence of Hot Jupiters on host stars with thick outer convection zones can occur, which either inhibits the typical spin-down of such stars, or has had an influence on the initial rotation of the stars when they arrive on the main sequence through changes in the star-disk coupling at young ages of the system. We are conducting observations of a larger sample of such systems to substantiate these initial findings.
   \end{enumerate}

\begin{acknowledgements}
      This work was performed in part under contract with the California Institute of Technology (Caltech) funded by NASA through the Sagan Fellowship Program executed by the NASA Exoplanet Science Institute.
\end{acknowledgements}

\bibliographystyle{aa}
\bibliography{/home/kpoppen/texmf/katjasbib.bib}

\end{document}